\documentclass[%
 reprint,
 amsmath,amssymb,
 aps,
]{revtex4-2}
\usepackage{float}

\usepackage{graphicx}
\usepackage{dcolumn}
\usepackage{bm}

\begin{document}


\title{Direct comparison of two spin squeezed optical clocks below the quantum projection noise limit}

\author{John M. Robinson}
\email{john.robinson@colorado.edu}
\author{Maya Miklos}%
\author{Yee Ming Tso}
\author{Colin J. Kennedy}
\author{Tobias Bothwell}
\author{Dhruv Kedar}
\author{James K. Thompson}
\author{Jun Ye}
\email{ye@jila.colorado.edu}

\affiliation{%
 JILA, National Institute of Standards and Technology and University of Colorado, Department of Physics, Boulder, Colorado 80309, USA
}%



\date{\today}

\begin{abstract}
Building scalable quantum systems that demonstrate genuine performance enhancement based on entanglement is a major scientific goal for fields including computing, networking, simulations, and metrology~\cite{Giovannetti_nat_photonics, Cirac_PRL, Polzik_PRA}.
The tremendous challenge arises from the fragility of entanglement in increasingly larger sized quantum systems.
Optical atomic clocks utilizing a large number of atoms have pushed the frontier of measurement science~\cite{Bothwell_nature, Oelker_nat_photonics, Mcgrew_Nature}, building on precise engineering of quantum states and control of atomic interactions~\cite{Aeppli_science_advances}. 
However, today’s state-of-the-art optical atomic clocks are limited by the quantum projection noise (QPN) defined by many uncorrelated atoms~\cite{Oelker_nat_photonics, Marti_PRL}. 
Pioneering work on producing spin squeezed states of atoms has shown a path towards integrating entanglement into the best performing clocks~\cite{Hosten_Nature, Edwin_Nature}. 
However, to directly demonstrate advantage of quantum entanglement in a working clock we must prevent backaction effects that degrade quantum coherence and introduce uncontrolled perturbations, as well as minimize the influence of technical noise arising from the interrogating clock laser. 
Here we present a new optical clock platform integrated with collective strong-coupling cavity QED for quantum non-demolition (QND) measurement.
Optimizing the competition between spin measurement precision and loss of coherence, we measure a Wineland parameter of -1.8(7) dB for 1.9$\times$ 10$^4$ atoms, thus verifying the presence of entanglement.
Furthermore, a moving lattice allows the cavity to individually address two independent sub-ensembles, enabling us to spin squeeze two clock ensembles successively and compare their performance.
This differential comparison between the two squeezed clocks directly verifies enhanced clock stability of 2.0(3) dB below QPN, and 0.6(3) dB above the standard quantum limit (SQL), at the measurement precision level of 10$^{-17}$, without subtracting any technical noise contributions. 

\end{abstract}

\maketitle




\maketitle
Optical atomic clocks are rapidly advancing the frontier of measurement science with continued progress in their precision and accuracy.
Accuracy evaluations at the $10^{-18}$ level~\cite{Bothwell_Metrologia, Brewer_PRL, Mcgrew_Nature} and frequency ratio measurements in networks of atomic clocks are setting the stage for the redefinition of the SI second based on optical technology~\cite{BACON_nature, Roberts_NJP}.
Besides time-keeping, advanced atomic clocks are also being employed for tests of fundamental symmetry and searches for new physics, as well as applications in relativistic geodesy and quantum information science \cite{Sanner_Nature, Kennedy_PRL, Takamoto_nat_photonics, Schine_nat_physics}.
Clock precision on the $21^{\text{st}}$ digit has recently enabled the measurement of the gravitational redshift within a single atomic ensemble at the sub-mm length scale~\cite{Bothwell_nature}.
Improving upon the fundamental limits of optical clock stability promises to open new opportunities in physics.

A fundamental noise source in atomic clocks is the quantum projection noise that stems from the inherent population fluctuations associated with the projective measurement of $N$ uncorrelated atoms~\cite{Itano_PRA}.
With QPN-limited stability scaling as $1/\sqrt{N}$, operating with a higher atom number $N$ is advantageous.
However, technical noise from imperfect state readout, intrinsic atom-atom interactions, or aliased frequency noise of the interrogating clock laser pose challenges for observing clock performance at the QPN limit~\cite{AlMasoudi_PRA}.
With precise engineering of quantum states and control of atomic interactions~\cite{Aeppli_science_advances}, and using laser-noise mitigation techniques such as synchronous comparisons~\cite{Marti_PRL,Oelker_nat_photonics}, state-of-the-art optical clocks are currently approaching QPN-limited stability with up to $10^5$ atoms~\cite{Bothwell_nature}. 

\begin{figure*}
\centerline{\includegraphics[scale=1.25]{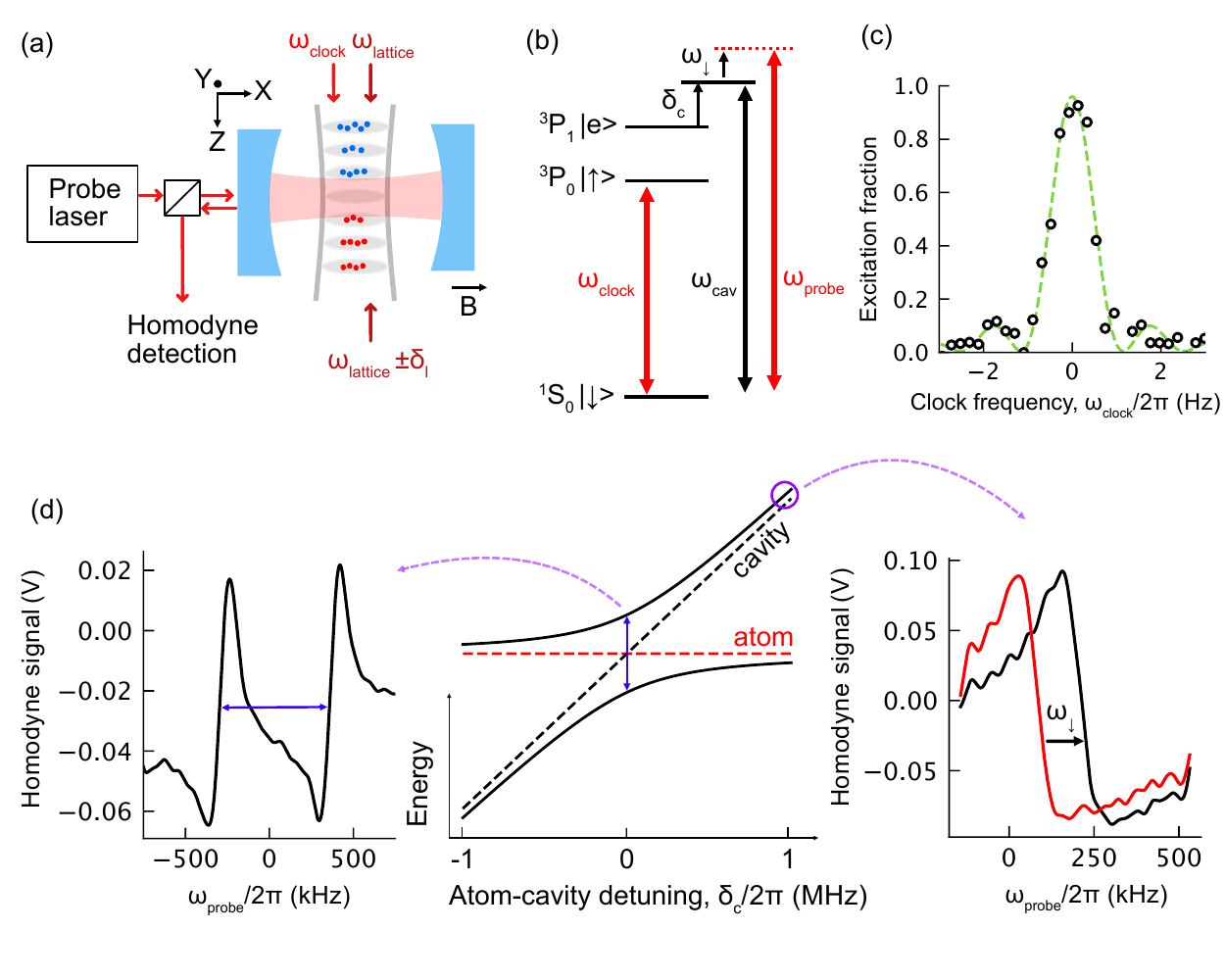} }
\caption{ \textbf{Optical clock with cQED architecture}
\textbf{(a)}$^{87}$Sr atoms are trapped in a movable vertical optical lattice (by detuning one lattice beam by $\delta_l$), enabling both independent squeezing and readout of two sub-ensembles (red and blue) within the atomic cloud.
The clock laser propagates along the vertical lattice, providing a global drive of the clock transition. 
Populations are measured non-destructively via homodyne detection of the laser probing the atom-cavity system.
\textbf{(b)} The relevant energy levels of $^{87}$Sr. We prepare coherent superpositions between the clock states $\mid\downarrow\rangle$ and $\mid\uparrow\rangle$. 
The frequency of the optical cavity $\omega_{cav}$ is tuned near the $\mid\downarrow\rangle\rightarrow \mid$$e\rangle$ transition in order to realize the atom-cavity coupled system. 
Atoms in the ground state $N_{\downarrow}$ shift the cavity frequency by an amount $\omega_{\downarrow}$, while atoms in $\mid\uparrow\rangle$, the other optical clock state, do not couple to the cavity. 
\textbf{(c)} Rabi spectroscopy of the clock transition with a Fourier-limited linewidth of 1~Hz using a $\pi$-pulse of 0.8~s. 
Open black circles indicate the measured data, with the corresponding Rabi fit as the dashed green line. 
\textbf{(d)} Left: The measured vacuum Rabi splitting indicates we are in the collective strong-coupling regime.
Middle: Avoided crossing behavior of the atom-cavity system. 
Right: When the cavity is slightly detuned by $\delta_c = 2\pi\times~1$ MHz we observe the dispersive shift of the cavity-like mode. }
\label{fig:1}
\end{figure*}

The development of quantum entanglement has provided an exciting new direction for reducing the impact of QPN in quantum sensors, offering the opportunity to greatly advance upon this state-of-the-art performance.
A particular form of entanglement, the spin squeezed state (SSS), was proposed early on to utilize quantum correlation to conceal noise from individual atoms and thus achieve improved measurement precision and bandwidth~\cite{Wineland_PRA,Kitagawa1993}.
The creation of entanglement for metrology has been explored in a wide variety of atomic quantum sensors including microwave clocks \cite{appel2009mesoscopic,Leroux2010a,Leroux2010b,SchleierSmith2010,Chen2011,Cox_PRL,Hosten_Nature, Pezze_2018}, ion clocks\cite{Nichol_nature,Marciniak_Nature}, magnetometers \cite{PolzikSqueezedMagnetometer2010}, and matterwave interferometers \cite{GreveLuo2022}. 

Spin-squeezing in atomic clocks has yet to yield enhancement at state-of-the-art stability levels. 
A spin-squeezed microwave clock has observed 11 dB of enhancement \cite{Hosten_Nature} at the $10^{-10}$ stability level at 1 s, in contrast with microwave fountain clocks at $10^{-14}$~\cite{Weyers_Metrologia}. 
For optical clocks, which operate at much higher stability, generation of entanglement has been demonstrated by a measured Wineland parameter of -4.6 dB ~\cite{Edwin_Nature}. 
After subtraction of a laser noise model, an optical clock employing a SSS was inferred to operate -4.4 dB below the SQL at a fractional frequency stability of $1.3\times10^{-13}~\tau^{-1/2}$ (where $\tau$ is the averaging time in seconds)~\cite{Edwin_Nature}.
However, metrological applications require direct observation of clock stability enhancement without post-processed removal of technical noise.
It remains to be seen whether spin-squeezing will enhance the operation of state-of-the-art atomic clocks. 

\begin{figure*}
\centerline{\includegraphics[scale = 1.1]{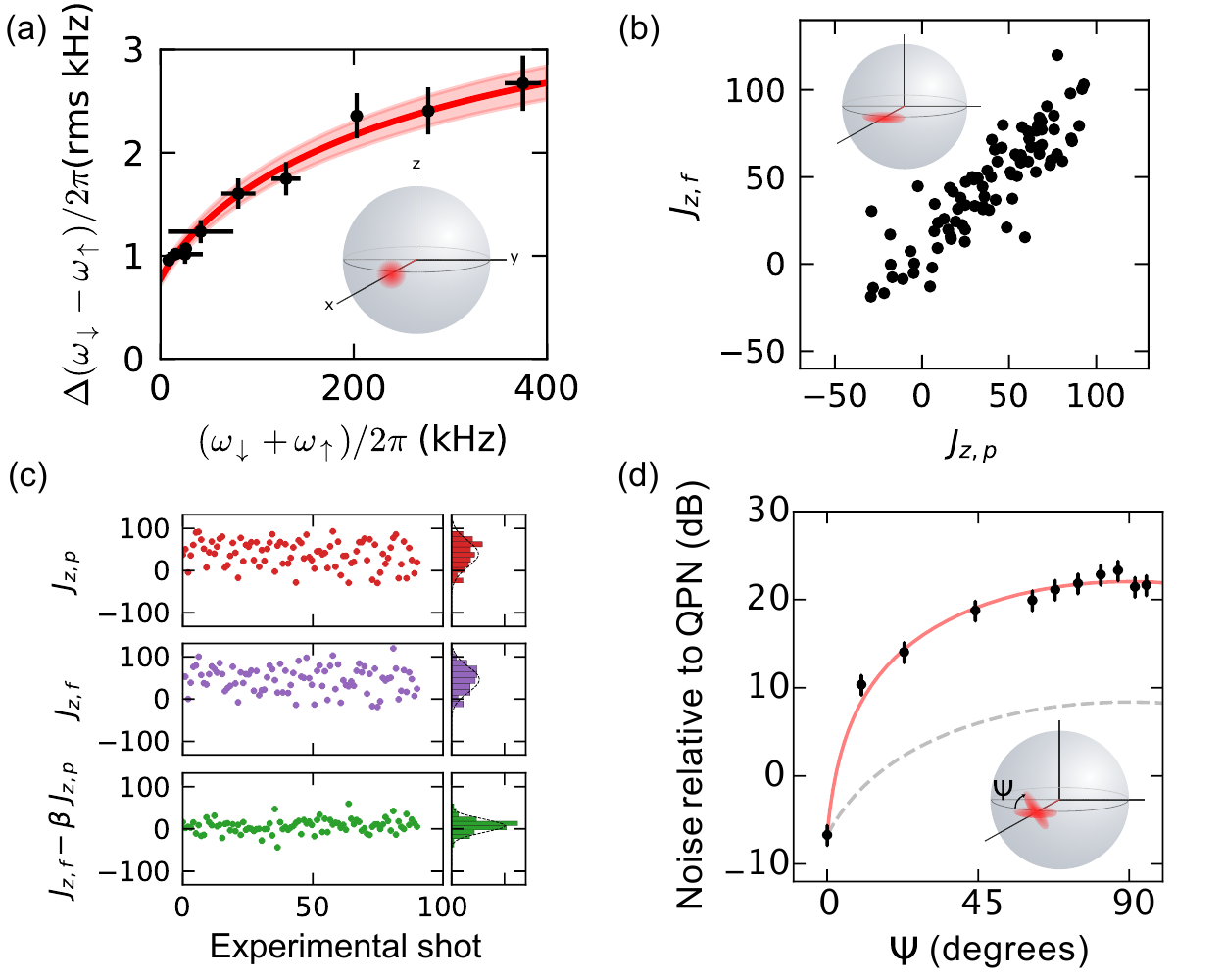}}
\caption{\textbf{Non-demolition measurements of the collective spin state} 
\textbf{(a)} Measurement of the QPN fluctuations of the initial CSS. Red line: fit to the data, giving  $g =2\pi\times 5.2(2)$ kHz, with the shaded area indicating the 1-sigma confidence interval. 
Inset: Pictoral representation of the CSS on a Bloch sphere. 
\textbf{(b)} The high-degree of correlations are shown between $J_{z,p}$ and $J_{z,f}$. Inset: SSS shown on the Bloch sphere.
\textbf{(c)} The pre measurements $J_{z,p}$ (red), final measurements $J_{z,f}$ (purple) and the difference $J_{z,f}-\beta$$J_{z,p}$ (green) are shown for each experimental shot.
\textbf{(d)} State tomography of the SSS. The measured noise relative to the QPN as a function of rotation angle $\psi$.
The solid red line is a fit to the measured data, and the dashed gray line indicates a unitary spin-squeezed state.}
\label{fig:2}
\end{figure*}

In this work, we report the design and operation of a spin-squeezed clock to specifically address these outstanding challenges. 
An optical lattice clock employing $10^4$ atoms is interrogated with a state-of-the-art optical local oscillator~\cite{Matei_PRL, Oelker_nat_photonics}, ensuring competitive performance for both atomic coherence and clock stability.
A collective, strongly-coupled cavity QED system is used to perform QND measurement of the clock state~\cite{Bowden_PRX}, providing spin squeezing and clock readout. 
We generate a SSS with a single ensemble of $N$ = 2.4$\times$10$^4$ atoms and directly measure a Wineland parameter of -1.8(7) dB, proving the creation of entanglement in this sample.
A movable optical lattice intersecting the cavity mode is used to transport atomic ensembles into and out of the cavity mode to address multiple independent clock ensembles.   
By alternately shuttling two spatially separated sub-ensembles in and out of the cavity, we demonstrate the first direct comparison between two spin-squeezed optical clocks at 2.0(3) dB below QPN, and 0.6(3) dB above the standard quantum limit. 
This comparison performs not far from the state-of-the-art, averaging down to the level of 10$^{-17}$ measurement precision, setting the stage for future entanglement-enhanced measurements that rival the top-performing classical clocks.

Our clock operates with up to 1.9$\times$10$^4$ $^{87}$Sr atoms confined in a vertical one-dimensional (1D) magic wavelength optical lattice (Fig \ref{fig:1}a).
The clock laser propagates along the vertical trapping lattice, globally addressing all of the atoms on the ultra-narrow $\mid\downarrow\rangle \equiv\mid ^1$$S_0,m_F = +9/2\rangle$ to $\mid\uparrow\rangle \equiv \mid^3$$P_0, m_F = +9/2\rangle$ clock transition (Fig. \ref{fig:1}b).
We demonstrate Rabi spectroscopy at a Fourier-limited full-width-half-maximum of 1 Hz with a peak $\pi$-pulse transfer efficiency of $97(1)\%$ (Fig. \ref{fig:1}c).
While our squeezed-clock experiments currently operate at shorter interrogation times, this demonstrates the capability of achieving state-of-the-art laser-atom coherence times. 

The key novel features of the system are the combination of the state-of-the-art clock spectroscopy with QND-based spin squeezing, and the capability of moving clocks with entangled atoms for direct comparison.
The coupled atom-cavity system is realized by tuning the bare optical cavity near the  $\mid\downarrow\rangle$$\rightarrow\mid$$e\rangle\equiv \mid^3$$P_1, m_F=+11/2\rangle$ transition~($\Gamma = 2 \pi \times 7.48(1)$~kHz)~\cite{NorciaPRA16}.
The effective vacuum Rabi frequency for atom-cavity coupling is $2g = 2 \times \left(2\pi \times 5.2(2)\right)$ kHz (Fig. \ref{fig:2}a, see Methods)~\cite{Hood_PRL,SchleierSmith2010, Cox_PRL, Hu_PRA}. 
With the bare cavity photon decay rate of $\kappa = 2\pi\times 158(7)$~kHz, we have a single-atom effective cooperativity of $\mathcal{C} = \frac{4g^2}{\kappa\Gamma} = 0.1$.
For effective atom number $N$ = 10$^4$ we are well into the collective strong-coupling regime with $N\mathcal{C}$=10$^3$, as is clearly seen by the vacuum Rabi splitting (Fig.~\ref{fig:1}d).

To optimize the information we gain about the collective spin state over the loss of coherence, we detune the cavity resonance from $\mid\downarrow\rangle\rightarrow\mid$$e\rangle$ by $\delta_c = 2\pi \times 1$ MHz (Fig. \ref{fig:1}b).
A fixed-frequency laser is then parked on resonance of the cavity-like mode to measure the dispersive shift $\omega_\downarrow$.
We express the number of atoms in $\mid\downarrow\rangle$ in terms of $\omega_{\downarrow}$ as $N_\downarrow = 
\omega_\downarrow\frac{\delta_c}{g^2}(1 + \frac{\omega_\downarrow}{\delta_c})$ (see Methods). 
An optical $\pi$-pulse is then applied to swap the population between   $\mid\uparrow\rangle$ and $\mid\downarrow\rangle$. 
The frequency shift is measured again to determine the excited state population $N_\uparrow$.
We generate entanglement using conditional spin squeezing by making repeated QND measurements of the collective spin projection $J_z = (N_{\downarrow}-N_{\uparrow})/2$~\cite{Chen2011}. 
Two repeated measurements of $J_z$ contain highly correlated QPN, hence their difference allows one to perform sub-QPN metrology~\cite{Leroux2010a, Chen2011}.

\begin{figure}
\centerline{\includegraphics[width=3.25 in]{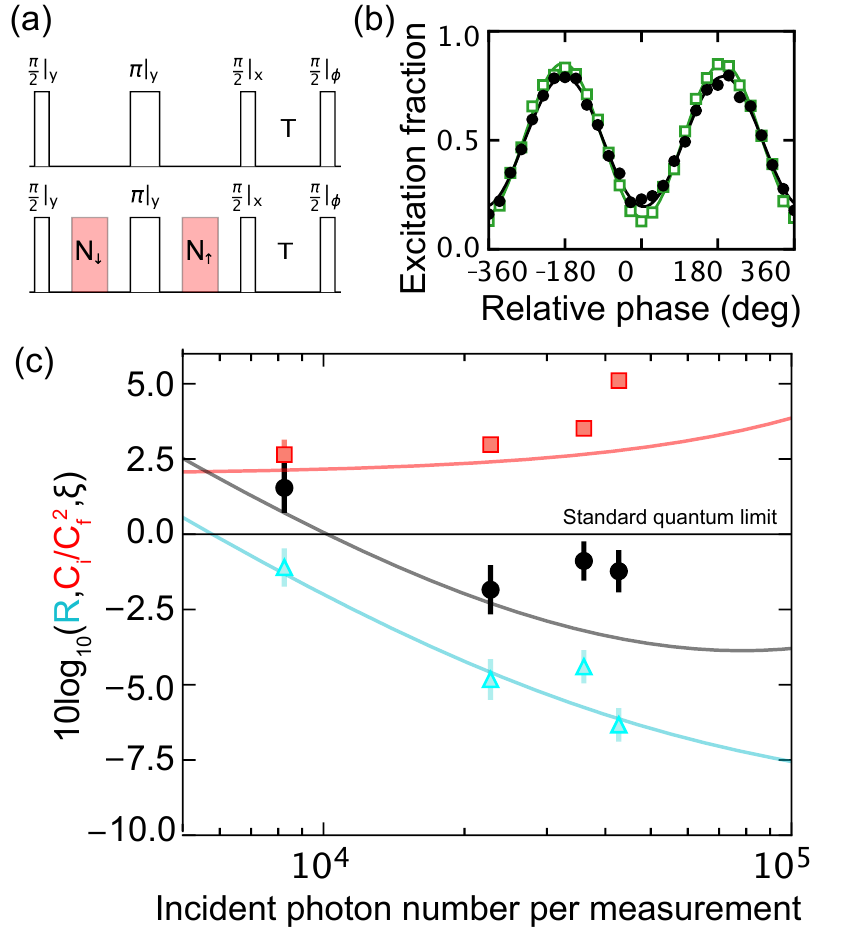}}
\caption{\textbf{Directly observed Wineland parameter versus QND probe strength} \textbf{(a)} The degree of atomic coherence is measured by scanning the Ramsey fringe with two distinct pulse sequences. Top: Sequence with all optical rotations required for squeezing but with no probing applied. Bottom: sequence with QND probing.
\textbf{(b)} Ramsey fringes for the two corresponding pulse sequences taken at a probe photon number of $2.3\times 10^4$ and a Ramsey dark time of $T = 14$~ms. The reduction of contrast at this probe power is 11(1)\%. 
\textbf{(c)} Relative spin noise reduction $R$ (cyan circles), fractional contrast loss $C_i/C_f^2$ (red circles), and the corresponding Wineland parameter $\xi$ (black circles). 
At the optimal photon number, we directly measure a Wineland parameter of -1.8(7) dB.
Expected $R$ given our estimated quantum efficiency of $Q = 0.28$ (cyan line), expected contrast loss via free-space scattering (red line), and the corresponding expected Wineland parameter (black line, see Methods).
}
\label{fig:3}
\end{figure}

To properly quantify spin squeezing, we first measure the QPN of a coherent spin state (CSS).
A $\pi/2$ pulse prepares the CSS on the equator of the Bloch sphere, and a measurement of $J_z$ is performed. 
The standard deviation of $J_z$, expressed in units of cavity frequency shift, is then fit to determine the effective atom-cavity coupling, extracting $g =2\pi\times 5.2(2)$~kHz (Fig. \ref{fig:2}a), compared to an independently estimated value of $g =2\pi\times 4.8(2)$ kHz based on cavity parameters and atomic cloud distributions (see Methods).

We now demonstrate repeated measurements of $J_z$ with differential resolution well below QPN. 
After an initial $\pi/2$, we make a pre-measurement denoted by $J_{z,p}$, wait a typical dwell time of 20 ms, and perform the final measurement $J_{z,f}$ (Fig. \ref{fig:2}b).
For the photon number of 2.3$\times 10^4$, the data was taken with a dwell time of 4 ms. 
We have found that the spin noise reduction does not depend on dwell time at these short times.
While the pre-measurement fluctuates with standard deviation $\Delta J_{z,QPN} = \sqrt{N}/2$, the final measurement contains highly correlated QPN, as seen by the reduced noise in the difference between the two measurements (Fig. \ref{fig:2}c).
Spin noise reduction is defined as $R = \left(\frac{\Delta (J_{z,f}-\beta~J_{z,p})}{\Delta J_{z,QPN}}\right)^2$, where we use an optimal estimator $\beta$ to account for differential technical noise.
We directly observe spin-noise reduction $R = -4.8(6)$ dB relative to QPN at the optimal squeezing photon number.
Assuming the detection noise between the two measurements is uncorrelated, we can infer the intrinsic spin-noise reduction of $-6.7(6)$ dB.
We perform state tomography to evaluate the amount of anti-squeezing introduced by the QND measurement (Fig. \ref{fig:2}d).
The observed anti-squeezing is well above the expected level given our estimated quantum efficiency of $Q = 0.28$, indicating additional technical noise in the anti-squeezing quadrature (see Methods).

To certify that the measured spin noise reduction arises from entanglement, one must weigh $R$ against the loss of coherence induced by the probe as measured by the Ramsey fringe contrast.
The Wineland parameter, serving as a criterion for the generation of entanglement, is expressed as \cite{Wineland_PRA,Cox_PRL} :
\begin{equation}
\label{eq:Wineland}
    \xi = \left(\frac{\Delta(J_{z,f} - \beta J_{z,p})}{\Delta J_{z,QPN}} \right)^2 \frac{C_i}{C_f^2},
\end{equation}
where $C_i$ and $C_f$ are the initial and final contrast, respectively.
$\xi = 1$ corresponds to the SQL.
Generating and using entanglement directly in the optical clock states requires performing several optical rotations that induces loss of Ramsey contrast due to single-particle motional effects in the 1D optical lattice, even without QND probing.
This is in comparison to the generation of entanglement in a ground-state manifold, where the use of microwave rotations typically leads to less degradation of the atomic coherence~\cite{Braverman_PRL, Edwin_Nature}.
To distinguish the effect of our QND probing on optical atomic coherence, we measure the Ramsey fringe contrast with and without probe, keeping all other rotations in the sequence (Fig. \ref{fig:3}a).
The initial contrast with no QND probing is $C_i=0.71(1)$ (Fig. \ref{fig:3}b).
Turning on a QND probe with $2.3\times10^4$ photons per population measurement reduces the contrast to $C_f=0.60(1)$. 
At this optimal probe power, the measured Wineland parameter reaches -1.8(7) dB (Fig. \ref{fig:3}c), a direct verification of spin entanglement in our system. 
Assuming uncorrelated detection noise in the two $J_z$ measurements, we infer a Wineland parameter of -3.7(7) dB.

\begin{figure}
\centerline{\includegraphics[scale=1]{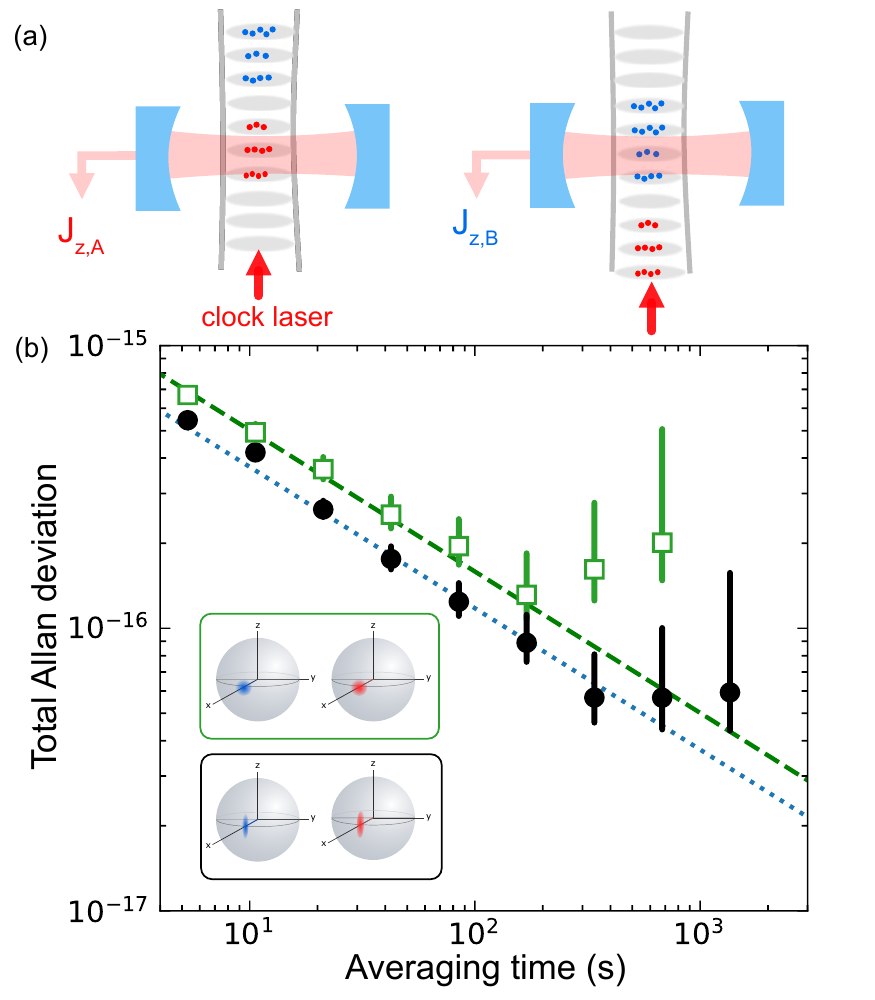} }
\caption{ \textbf{Differential clock comparison} \textbf{(a)} The optical lattice is moved in order to shuttle the two sub-ensembles in and out of the cavity, allowing for independent squeezing and readout. The clock laser, coming from below, address both ensembles in a global fashion. \textbf{(b)} Allan deviation of the CSS-CSS comparison (green open squares) and of the SSS-SSS comparison (black circles), showing an enhancement of 2.0(2) dB using spin squeezing. 
Theoretical QPN limit of the clock comparison (dashed green line), and the theoretical SQL limit (dotted blue line). The inset shows the different states on the Bloch sphere for each clock comparison.   }
\label{fig:4}
\end{figure}
To demonstrate the spin squeezing gain on clock performance, we perform a direct clock comparison between two spin-squeezed ensembles.
Using the moving optical lattice, we address two independent sub-ensembles labelled A and B (Fig. \ref{fig:4}a) within the same atomic cloud, separated by a vertical distance of 150 $\mu$m  (see Methods).
We perform differential clock comparisons between the two sub-ensembles, contrasting the case where both are projected into SSSs (black circles) against the case of CSSs without the use of the QND probe (green open squares) (Fig. 4b). 
In either case, all rotations and transports that manipulate the states are performed identically, allowing for a direct measurement of the impact of spin squeezing on clock stability.
The observed stability is $1.58(3)\times 10^{-15}~\tau^{-1/2}$ and $1.25(2)\times 10^{-15}~\tau^{-1/2}$ for the CSS - CSS and SSS - SSS comparison, respectively. 
We directly observe an enhancement of stability by 2.0(2) dB in the SSS - SSS comparison over that of CSS - CSS.

To put our results in the proper context, we seek to benchmark the observed SSS-SSS stability to both the practically achievable limit set by QPN and fundamental limit set by SQL. 
The measurement of $(J_{z,A} - J_{z,B})$ is limited by the quadrature sum of QPN arising from each sub-ensemble.
The QPN-limit is then $1/\left(C_i\sqrt{N_A + N_B}\right)$, using $C_i$ as the slope of the Ramsey fringe to convert QPN into fractional frequency noise.
With a measured $C_i=0.55(1)$ for both ensembles, this sets the bound of optimal practically achievable stability for CSS - CSS (dashed green line, Fig. \ref{fig:4}b). 
However, the ultimate bound on the performance of an unentangled ensemble is the SQL. 
This strictest bound treats the $(1-C_i)$ fraction of atoms as no longer participating in the pure CSS, thus reducing projection noise to give the SQL-limited stability $1/\sqrt{C_i (N_A+N_B)}$ (dotted blue line, Fig. \ref{fig:4}b).

The observed stability of the CSS - CSS comparison is consistent with the QPN-limited stability, and 2.6(3) dB above SQL. 
Implementing QND-based squeezing operation and accounting for the final contrast of $C_f$ = 0.50(1), the SSS - SSS comparison shows a 2.0(3) dB gain over the QPN-limited stability, demonstrating practical metrological enhancement from the squeezing operation.
This result is above SQL by 0.6(3) dB, indicating a Wineland parameter near unity.

The direct observation of the clock comparison below the QPN limit with measurement precision averaging down to the $10^{-17}$ level is a crucial step towards improving the performance of the best optical lattice clocks via entanglement.
Such improvement in differential clock stability translates directly into increased sensitivity for many applications of interest, including the measurement of the gravitational redshift at ever-shorter length scale and the future development of clock networks for fundamental physics~\cite{Komar_nature_physics,Nichol_nature, Roberts_NJP, Kolkowitz_PRD}.
Further improvement will come from enhanced control of atomic motion, which will yield improved spin rotation fidelity and increased coupling to the cavity mode.
Larger atom number, improved single-atom cooperativity, and better quantum efficiency will all lead to stronger spin squeezing.
By integrating the exquisite stability of a competitive optical lattice clock with the all-to-all interactions enabled by cavity QED, this system opens the door for explorations of other flavors of entanglement-enhanced metrology~\cite{Davis_PRL, Kaubruegger_PRX, Colombo_nat_physics}, as well as implements important tunability and flexibility for studies of a wide variety of many-body spin dynamics~\cite{Zhang_science, Bromley_nat_physics, Muniz_nature, Vaidya_PRX, Davis_PRL_2020, Dogra_science}.


\bibliography{apssamp}

\section*{Methods}
\subsection*{Atomic state preparation}
$^{87}$Sr atoms are laser-cooled and trapped in a two-stage magneto-optical trap located $\approx 40$ mm below the cavity, where the atoms are subsequently loaded into the vertical optical lattice formed by two counter-propagating 813 nm beams.
The relative phase of the optical lattice is detected by interfering the two beams in a Mach-Zender interferometer, and we stabilize the optical lattice phase by feeding back to an acousto-optic modulator (AOM) on the bottom-up lattice beam with $\approx$ $1$ kHz bandwidth.

Once cooled and trapped in the lattice, the atoms are optically pumped into the $\mid$$m_F$=+9/2$ \rangle$ hyperfine ground state. 
To transport the atoms vertically into the cavity, the frequency of the bottom-up lattice beam is linearly ramped and detuned by $\delta_{l}$, resulting a moving lattice with a velocity of $v(t) = \frac{\delta_{l}(t) \lambda_{l}}{4\pi}$ \cite{sauer2004cavity}. 
To select the atoms with the lowest temperature, we apply a ramp of the optical lattice trap depth down to 7$E_r$ and then back up to our operation depth of 20$E_r$, allowing the hot atoms to escape the trap.
We apply a bias magnetic field of 2 $G$ along the direction of the cavity (Fig. \ref{fig:1}a).
To ensure high spin state purity of our atoms, a clock $\pi$-pulse of 40 ms is applied on the $\mid^1$$S_0,m_F=+9/2\rangle$$\rightarrow$$\mid^3$$P_0,m_F=+9/2\rangle$ transition, and the atoms remaining in the ground state are removed by applying a light pulse tuned to the $\mid^1$$S_0\rangle$$\rightarrow$$\mid^1$$P_1 \rangle$ 461 nm transition. 
After these preparation steps we have an ensemble of atoms with the radial temperature of 290(10) nK and vertical cloud size of 130 $\mu$m. 

\subsection*{Optical local oscillator}
The clock local oscillator is a 698 nm fiber laser pre-stabilized by a 40 cm ultralow expansion glass cavity.
This 698 nm laser is phase locked to a frequency comb that is stabilized by a 21~cm crystalline silicon cavity operating at 124~K~\cite{Matei_PRL,Oelker_nat_photonics}. 
A portion of the fully stabilized 698 nm light seeds an injection locked laser, which is then delivered to the experiment via a fiber-noise cancelled optical fiber.
The $1^{st}$ order of an AOM is used to probe the atomic resonance, and the zeroth order of the AOM serves as the phase reference for the fiber noise cancellation.
Differential noise between the $0^{th}$ and $1^{st}$ order is minimized using beam tubes and multiple layers of isolation from acoustic noise.

\subsection*{Optical cavity and QND probe}
The optical cavity with cavity length of $L= 6.9720(2)$~cm supports a TEM00 mode with a $1/e^2$ beam waist of $w_0=71~\mu$m and has a power-decay rate of $\kappa=2 \pi \times 158(7)$ kHz. 
The cavity is one-sided such that the transmission coupling rate of the input mirror $\kappa_{1}$ is much greater than that of the back mirror, $\kappa_{2}$.
The cavity is isolated from vibrations by suspending the spacer in a double-pendulum configuration, using Viton as the lossy springs.
The vibration isolation results in a roll-off in vibration noise coupling above $\approx 30$~Hz.

We stabilize the cavity length to a pre-stabilized 813 nm laser via  Pound-Drever-Hall (PDH) locking, feeding back to the cavity PZT with a bandwidth of $\approx$ 1 kHz.
The 813 nm ECDL is pre-stabilized via a phase lock to the same frequency comb that transfers the clock local oscillator phase. 
For the QND probe light, an ECDL at 689 nm is stabilized to a Hz-level optical cavity, and at low frequency, stabilized to the frequency comb. 
A phase-modulated sideband at 137.59 MHz is then generated by a fiber electro-optic modulator (EOM) to probe the cavity. 
We set the probe photon number by changing the modulation depth of the EOM drive.
The technical noise floor of the entire locking chain is evaluated by probing the empty cavity at a probe photon number of $\approx$$10^6$ per measurement window. 
The standard deviation of two repeated bare cavity frequency measurement yields 200 Hz, approximately 20 dB below the QPN limit. 
Atoms trapped in the vertical 1D optical lattice have longitudinal trap frequency of 25 kHz and radial trap frequency 34 Hz.
The duration of each QND measurement is chosen to be 40 ms to average the single-particle motional effects of atoms. 

\subsection*{Balanced homodyne detection}
We measure the phase shift of the probe laser in the reflection port of the cavity using homodyne detection against a reference LO (Fig. \ref{fig:1}a).
The homodyne fringe is detected using a home built balanced photodetector, where we take the difference between the two output ports of the homodyne beamsplitter.
The combination of active stabilization of the LO intensity and the common-mode rejection of the LO intensity noise allows the homodyne detection to be photon shot noise limited.
The technical dark noise of the photodetector is $\approx~31$ dB below the LO photon shot noise. 
We stabilize the phase of the carrier with respect to the LO by detecting the carrier-LO beat note and phase locking it to a reference RF synthesizer.
By stabilizing this phase, we remove any path length fluctuations that arise in the differential path between the probe beam and the LO.
The power in the carrier is approximately 150 nW, so the photon shot noise of this phase lock is negligible compared to the probe sideband.

Vibrations of the optical breadboard couple to the homodyne output voltage via differential pointing instabilities of the LO beam through the interferometer.
An accelerometer placed on the table near the homodyne interferometer shows strong correlations with the homodyne output voltage at $\approx$$20$ Hz and $\approx$$30$ Hz, coming from air conditioning motor vibrations.
We simultaneously record the homodyne signal and the accelerometer output on an oscilloscope, and perform subtraction of the vibrations in order to reach the photon shot noise limit.
\subsection{Effective atom-cavity coupling}
To fit $g$ to our measured noise of the differential cavity frequency shift $\Delta(\omega_{\downarrow}-\omega_{\uparrow})$, we require an expression for this noise in terms of the sum shift $\omega_{sum} = \omega_{\downarrow}+\omega_{\uparrow}$ (Fig. \ref{fig:2}a).
The eigenvalues for the $\mid\downarrow\rangle$ and $\mid\uparrow\rangle$ states are
\begin{equation}
    \omega_{\downarrow,\uparrow} = \frac{\delta_c + \sqrt{\delta^2_c + \Omega^2_{\downarrow,\uparrow} }}{2},
\end{equation}
where the vacuum Rabi splitting for each spin state are $\Omega_\downarrow = 2g\sqrt{N_\downarrow}$ and $\Omega_\uparrow = 2g\sqrt{N_\uparrow}$.
The QPN fluctuation of the frequency shift is obtained using the derivative, $\Delta \omega_\downarrow = \lvert \frac{d\omega_\downarrow}{dN_\downarrow}\lvert \Delta N_\downarrow$.
For a CSS of $N$ total atoms prepared by a $\pi/2$ pulse on the equator of the Bloch sphere with state population $N_\downarrow =  N_\uparrow = N/2$, one can express $N$ in terms of the sum of measured frequency shifts $\omega_{sum} = \omega_\downarrow + \omega_\uparrow$ 
\begin{equation} \label{eq:atom_num}
    N = \omega_{\text{sum}}\frac{\delta_c}{g^2}\left(1+\frac{1}{2}\frac{\omega_{\text{sum} }}{\delta_c} \right).
\end{equation}
QPN for the two spin states is $\Delta N_\downarrow = \Delta N_\uparrow = \sqrt{N}/2$.
We can calculate the projection noise fluctuations of the frequency shift using the derivative of the eigenvalue expression.
Projection noise fluctuations of $N_{\downarrow}$ and $N_{\uparrow}$ are perfectly anti-correlated, and therefore $\Delta( \omega_\uparrow - \omega_\downarrow)$ is twice of the fluctuations of $\omega_\downarrow$, 
\begin{equation} \label{eq:std_dev_delta}
    \Delta( \omega_\uparrow - \omega_\downarrow)  = 2\Delta \omega_\downarrow =  \frac{g^2\sqrt{N}}{\sqrt{\delta_c^2 + \Omega_\downarrow^2}}.
\end{equation}
Using Eq.(\ref{eq:atom_num}) and Eq.(\ref{eq:std_dev_delta}), the expression for characterizing $g$ based on the measurement on QPN fluctuations of the initial CSS is
\begin{equation}
\label{eq:fit}
    \Delta( \omega_\uparrow - \omega_\downarrow) = g\sqrt{\frac{\omega_{sum}^2/2 + \delta_c \omega_{sum}}{(\omega_{sum} + \delta_c)^2}}
\end{equation}
To account for technical measurement noise in the absence of atoms ($\omega_{sum}=0)$, we fit Eq \ref{eq:fit} with an offset term added in quadrature.
At high atom numbers, rotation noise becomes noticeable in the QPN measurement.
By including a linear in $\omega_{sum}$ term in the fit equation, we can model rotation noise present in the measurement.
This result is consistent with subtracting independently measured rotation noise.
Our final result is a value of $g = 2\pi\times 5.2(2)$ kHz as shown in Fig~\ref{fig:2}a, and the bare cavity noise offset of $2\pi\times 0.76(5)$ kHz.

The effective atom-cavity coupling $g$ is independently estimated as a consistency check on our experimentally determined value from Fig. \ref{fig:2}a. 
We follow the convention from \cite{Hu_PRA}, where the effective $g$ is
\begin{equation}
\label{eq:geff}
    g^2 = \frac{\langle g_i^4 \rangle}{\langle g_i^2 \rangle},
\end{equation}
where $g_i$ is the atom-cavity coupling for the i$^{\text{th}}$ atom.
The effective atom number is 
\begin{equation}
    N = N_{\text{tot}}\frac{\langle g_i^2 \rangle^2}{\langle g_i^4 \rangle}
\end{equation}
where $N_{\text{tot}}$ is the total atom number.
In this work, when we refer to $g$ and $N$, we refer to the effective quantities.
Our coordinate system is defined such that X is the direction along the cavity axis, Y is the other horizontal axis orthogonal to X, and Z is the vertical direction along gravity.
The atomic density distribution in Z is modeled as a Gaussian, $\rho_{\text{Z}}(\text{Z}) = \frac{N}{\sqrt{2\pi}\sigma_{\text{Z}}}e^{-\text{Z}^2/(2\sigma_{\text{Z} }^2)}$, with standard deviation $\sigma_Z$. 
The probability distribution along Y is $P_{\text{Y}} (\text{Y}) = \frac{1}{\sqrt{2\pi}\sigma_{\text{Y} }}e^{-\text{Y}^2/(2\sigma_{\text{Y} }^2)}$ is a Gaussian with a standard deviation $\sigma_{\hat{\text{y}}}$ set by the thermal cloud radius.
We calculate $\sigma_{\text{Y}}$ from the radial temperature of $T_r = 290(10)$ nK from a radial Doppler scan of the clock transition and the radial trap frequency of $34(3)$~Hz.
If we allow the atoms to sufficiently time-average the standing wave along $x$, we have 
\begin{equation}
    g_i^2(\text{Y},\text{Z}) = \frac{g_0^2}{2} e^{-2(\text{Y}^2+\text{Z}^2)/w_0^2}
\end{equation}
with peak coupling $g_0 = d_{0}\sqrt{\frac{\omega_p}{2\epsilon_0 \hbar V}} =2\pi\times 8.6$ kHz. Here, $d_0$ is the dipole matrix element and $\omega_p$ is the angular frequency of the $\mid\downarrow\rangle\rightarrow\mid$$e\rangle$ transition, and $V = \frac{1}{4}\pi \omega_0 L$ is the effective cavity mode volume.
The ensemble averages are evaluated as
\begin{equation}
    \langle g_i^2 \rangle = \frac{1}{N} \int g_i^2(\text{Y},\text{Z}) \rho_{\text{Z}}(\text{Z})P_{\text{Y}} (\text{Y})d\text{Y}d\text{Z}
\end{equation}
and
\begin{equation}
    \langle g_i^4 \rangle = \frac{1}{N} \int g_i^4(\text{Y},\text{Z}) \rho_{\text{Z}}(\text{Z})P_{\text{Y}}(\text{Y})d\text{Y}d\text{Z},
\end{equation}
and when combined using Eq. \ref{eq:geff} gives the estimated value of $g = 2\pi\times 4.8(2)$ kHz.

\subsection*{Quantum efficiency}
The overall quantum efficiency of the measurement plays a key role in QND-based spin squeezing.
The amount of attainable spin-squeezing is linearly proportional to the quantum efficiency~\cite{Chen_PRA}.
Furthermore, $Q<1$ leads to excess noise in the anti-squeezed quadrature.
We first estimate the overall quantum efficiency by considering the different contributions, including the cavity quantum efficiency $\kappa_1/\kappa$ = 0.68, mode overlap of the cavity leakage light and the homodyne LO beam of 0.75, quantum efficiency of the photodiode of 0.88, and finite path efficiency of 0.62, and other negligible sources.
Multiplying these together gives a quantum efficiency of $Q = 0.28$.
We use this value when estimating the expected $R$ (cyan line in Fig. \ref{fig:3}c), showing reasonable agreement. The measured noise level at the anti-squeezed quadrature is 9 dB above what is expected from the estimated quantum efficiency (Fig.~\ref{fig:2}d). This indicates an additional noise source contributing to the anti-squeezing, but it does not preclude observing the benefit of spin squeezing.

\subsection*{Independence of the atomic sub-ensembles}
To assess the independence of the sub-ensembles, we vary the vertical separation between the ensembles and evaluate the Pearson correlation coefficient between the measured $J_z$ for each ensemble.
For no separation, we probe the same ensemble and expect high correlations between the measured QPN, and as we increase the separation distance, the correlation coefficient will decrease.
We first measure $J_{z,A}$ for ensemble A, apply a vertical displacement of the cloud via the movable optical lattice to put ensemble B in the cavity, and subsequently measure $J_{z,B}$.
The measured correlation coefficient between $J_{z,A}$ and $J_{z,B}$ is shown versus the separation distance between the ensembles in units of the mode waist w$_0$ (Extended Data Fig. \ref{fig:corr_coef}a).
The blue line is a numerical Monte Carlo simulation of the correlation coefficient for two ensembles with varying mean separation.
We also calculate the change of the combined QPN versus the separation distance using both analytical and numerical methods (Extended data Fig. \ref{fig:corr_coef}b).
For the differential clock comparisons of Fig. \ref{fig:4}, a spatial separation of 150~$\mu$m (dashed grey line) is chosen to guarantee independence of the ensembles.

\subsection*{Differential clock comparison}
The timing sequence for the differential clock comparison includes clock rotations, transports, and QND measurements (Extended data Fig. \ref{fig:SSS_SSS_pulse_sequence}a).
Clock pulses are shown as black pulses, with the pulse area and axis of rotation shown.
Transport steps are indicated by the green and purple pulses, and the transport waveform for the optical lattice detuning are linear ramps of the frequency over 5 ms.
All clock pulses are applied with the lattice at the same vertical location, so we do not have to take into account the varying clock laser phase.
After the pre-measurements, the $\frac{\pi}{2}$$\mid_{x}$ clock pulse rotates the spin-squeezed axis to the phase-sensitive axis.
After a total evolution time of $T = 14$ ms, the final $\pi/2$ pulse rotates back to the $J_z$-basis for readout.
The final measurements are taken at a higher probe photon number than the pre measurements.
The CSS - CSS comparison uses the same pulse sequence, but the pre-measurements have no probe light applied.
We set the relative separation between the two ensembles such that they give the same shift of $\approx 215$~kHz, corresponding to $\approx 8500$ atoms per ensemble.

The pre and final measurements for each ensemble constitute cavity frequency shifts for each spin state.
Making use of the eigenvalue expression for the atom-cavity system, we  convert these frequency differences to atom number differences, labelled $dN = N_{\downarrow}-N_{\uparrow}$.
To convert measured atom number differences directly
to phase, we scan out the Ramsey fringe (Extended data \ref{fig:SSS_SSS_pulse_sequence}b).
The fitted amplitude of this fringe $\alpha$ is used to convert the measured atom number difference for each ensemble to differential phase (in the small angle limit), 
\begin{equation}
\label{eq:phase_diff}
\phi_A - \phi_B = \frac{(dN_{A,f} - \beta_A dN_{A,p}) - \beta_D(dN_{B,f} - \beta_B dN_{B,p})}{\alpha}
\end{equation}
where we have introduced optimal estimators $\beta_A, \beta_B$ for the pre-measurements and $\beta_D$ for the differential noise of ensemble A and B.
The three parameters are simultaneously optimized to give the smallest $\Delta(\phi_A - \phi_B)$.
By varying the length of the dataset from half the length to the full length, we take the mean and standard deviation of each optimal estimator.
The values are $\beta_A = 0.49(1)$, $\beta_B = 0.48(1)$ and $\beta_D = 0.907(5)$ (Extended data \ref{fig:SSS_SSS_pulse_sequence}c,\ref{fig:SSS_SSS_pulse_sequence}d).
The slight deviation of $\beta_D$ from unity indicates some small asymmetric noise between the ensembles.
This could arise from inhomogenous ac Stark shifts, unequal squeezing of the ensembles, or differential thermal motion impacting clock rotations.
This $\beta_D$ is accounted for in the estimated SQL for the clock comparison,
\begin{equation}
\label{eq:SQL_diff_clock}
    \Delta\left(\phi_A - \phi_B\right)_{SQL} =\sqrt{ \frac{1}{C_i N_A} + \beta_D^2\frac{1}{C_i N_B}}.
\end{equation}

\section*{Acknowledgments}
We acknowledge funding support from National Science Foundation QLCI OMA-2016244, Department of Energy National Quantum Information Science Research Center - Quantum Systems Accelerator, Air Force Office for Scientific Research, DARPA, Vannevar Bush Faculty Fellowship, National Institute of Standards and Technology, and National Science Foundation Phys-1734006. M.M. acknowledges NSF Graduate Research Fellowship. We gratefully acknowledge early technical contributions and discussions from J. Meyer, J. Uhrich and E. Oelker. We acknowledge A. Aeppli, K. Kim, C. Sanner, R. Hutson, W. Milner, and L. Yan for many stimulating discussions. We thank A. M. Rey, C. Luo, E. Polzik, V. Vuleti\`c, J. Hall, M. Schleier-Smith and M. Kasevich for careful reading of the manuscript. 

\section*{Competing interests}
The authors declare no competing interests.
\section*{Contributions}
All authors contributed to the design and operation of the experiment, and data analysis and writing of the manuscript. 

\section*{Extended data figures}

\setcounter{figure}{0}
\begin{figure}[H]
\renewcommand\figurename{Extended data Fig.}
\centerline{\includegraphics[scale=0.5]{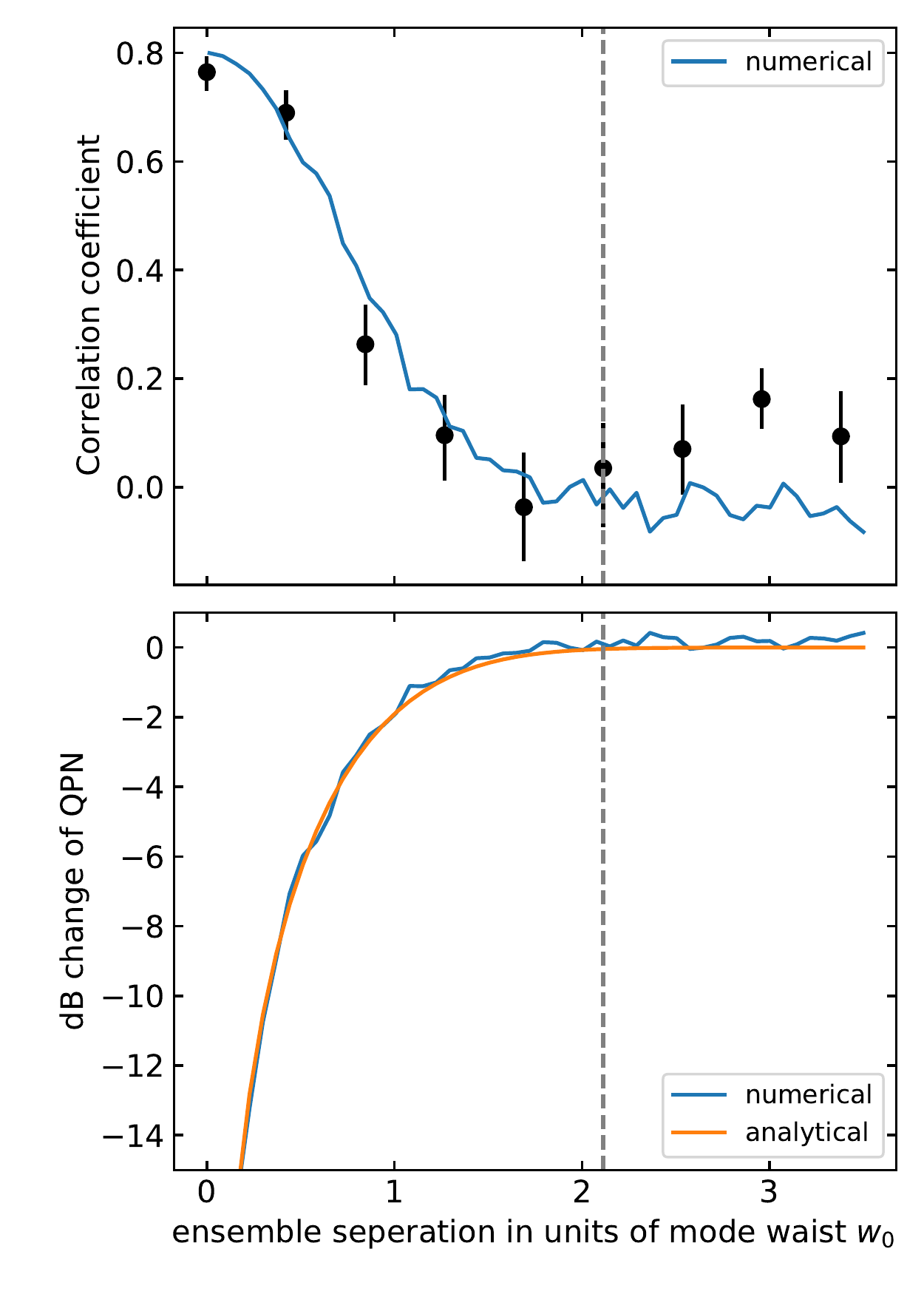} }
\caption{ \textbf{Independence of atomic ensembles} \textbf{(a)} Measured correlation coefficient between $J_{z,A}$ and $J_{z,B}$ versus the separation between the ensembles. (black circles). The blue line is a Monte Carlo simulation. \textbf{(b)} Corresponding change of the QPN due to the finite overlap of the ensembles, with numerical Monte Carlo simulation (blue) and analytical calculation (orange). At our operating ensemble separation, the change to QPN is 0.05 dB.}
\label{fig:corr_coef}
\end{figure}

\begin{figure*}
\renewcommand\figurename{Extended data Fig.}
\centerline{\includegraphics[scale=1.25]{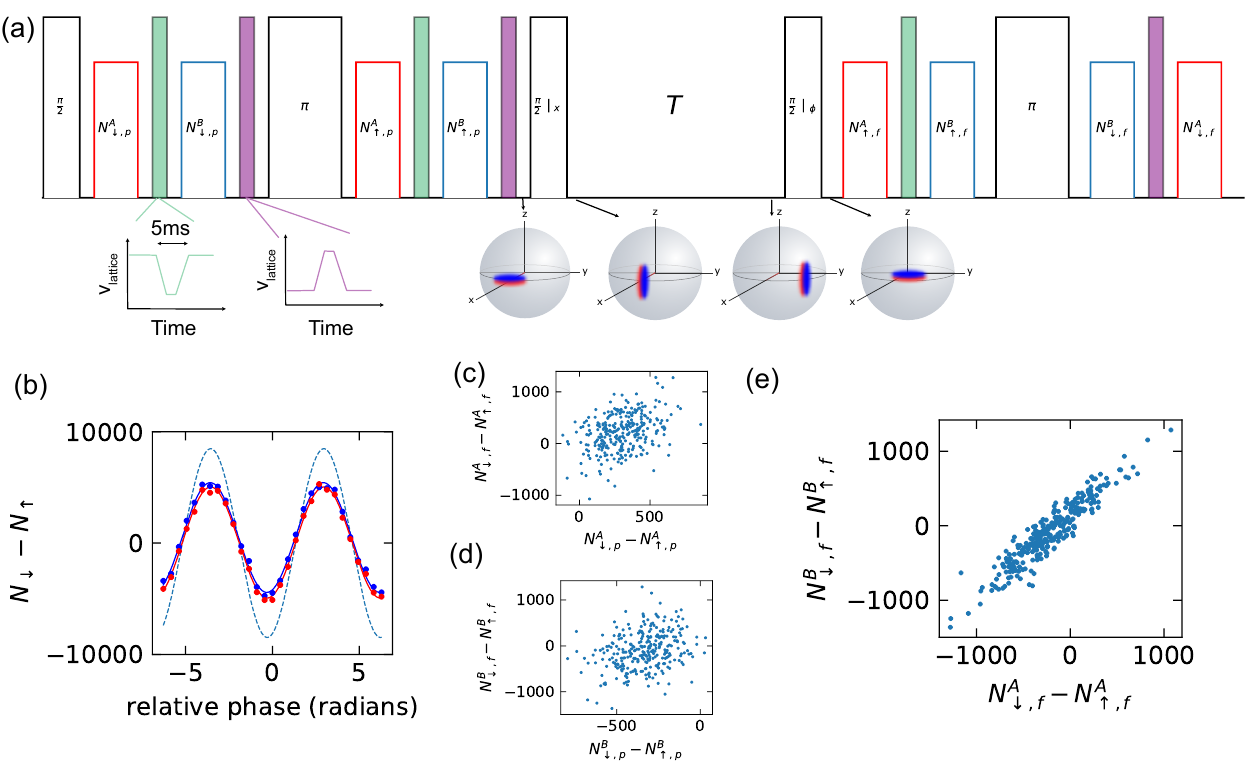} }
\caption{\label{fig:SSS_SSS_pulse_sequence} \textbf{Pulse sequence for SSS - SSS comparison} Clock pulses are the black pulses, measurements of ensemble A are the red pulses, and the transports are shown as the green and purple pulses. The Bloch spheres depict the spin state distribution at various points during the sequence. \textbf{(b)} Ramsey fringe measured by varying the phase of the final $\pi/2$ pulse. \textbf{(c)} Pre and final measurements of ensemble A. \textbf{(d)} Pre and final measurements of ensemble B. \textbf{(e)} The final measurements of ensemble A and B show strong correlations, allowing for the subtraction of the common-mode laser phase noise.}

\end{figure*}
\end{document}